# High Performance MoS$_2$ Phototransistors Photogated by PN Junction

*Seyed Saleh Mousavi Khaleghi[1,2,3], Jianyong Wei[1], Yumeng Liu[1], Zhengfang Fan[1], Kai Li[1], Kenneth B. Crozier[2,3,4], Yaping Dan[1, \*]*

[1]University of Michigan-Shanghai Jiao Tong University Joint Institute, Shanghai Jiao Tong University, Shanghai 200240, China

[2]Department of Electrical and Electronic Engineering, University of Melbourne, Victoria 3010, Australia

[3]ARC Centre of Excellence for Transformative Meta-Optical Systems (TMOS), University of Melbourne, Victoria 3010, Australia

[4]School of Physics, University of Melbourne, Victoria 3010, Australia

KEYWORDS: MoS$_2$ photodetector, photogating effect, Transition metal dichalcogenides (TMDs), Analytical photoresponse

ABSTRACT

Photodetectors based on two-dimensional (2D) atomically thin semiconductors suffer from low light absorption, limiting their potential for practical applications. In this work, we demonstrate a high-performance MoS$_2$ phototransistors by integrating few-layer MoS$_2$ on a PN junction formed in a silicon (Si) substrate. The photovoltage created in the PN junction under light illumination electrically gates the MoS$_2$ channel, creating a strong photoresponse in MoS$_2$. We present an analytical model for the photoresponse of our device and show that it is in good agreement with measured experimental photocurrent in MoS$_2$ and photovoltage in the Si PN junction. This device structure separates light absorption and electrical response functions, which provides us an opportunity to design new types of photodetectors. For example, incorporating ferroelectric materials into the gate structure can produce a negative capacitance that boosts gate voltage, enabling low power, high sensitivity phototransistor; this, combined with separating light absorption and electrical functions, enables advanced high-performance photodetectors.

There exists considerable interest currently for two dimensional (2D) semiconductors such as transition metal dichalcogenides (TMDs) and black phosphorus (bP) [1], [2], [3], [4] due to a wide range of potential applications in integrated circuits [5], [6], flexible electronics [7] and others [8]. The use of 2D materials in photodetection is particularly interesting and has attracted much attention in recent years [9], [10]. However, the ultra-thin nature of 2D materials means that light absorption is reduced correspondingly [11]. As a result, the photoresponses of 2D material photodetectors are often rather weak [12], [13]. An interesting strategy to enhance the photoresponses of these photodetectors is to introduce a relatively thick layer of quantum dots (QDs) onto the 2D material. High photoresponses have indeed been observed in these systems[14]. However, the operating principle (device physics) of these hybrid photodetectors remains elusive [15] and the process to deposit QDs is incompatible with complementary metal-oxide-semiconductor (CMOS) technology, making it difficult for device design and commercialization.

In this work, we demonstrated a MoS$_2$ phototransistor in which a PN junction in the substrate acts as the photogate. Light illumination beyond the absorption edge of MoS$_2$ creates a photovoltage in the PN junction that electrically gates the MoS$_2$ channel, thus inducing a high photogain in the MoS$_2$ channel. We propose an analytical device model for the photoresponse and find good agreement between it and the measured photocurrent. The work presented in this Letter offers a powerful new paradigm for designing high performance photodetectors based on 2D materials. By separating the light absorption and electrical response functions into distinct components, we can unlock unprecedented device performance. The light absorption is handled by a PN junction in the substrate, which can be engineered to maximize the optical absorption cross-section. The electrical readout is then performed by a high-transconductance MoS$_2$ channel, which can convert the absorbed photocurrent into a large, low noise electrical signal. This vertical integration allows the device footprint to remain extremely compact, enabling dense integration and scalability. The key advantage of this approach is that it allows us to independently optimize the optical and electronic properties of the device. We can maximize the light absorption and photocurrent generation in the PN junction, while simultaneously taking advantage of the exceptional electrical characteristics of the MoS$_2$ channel to achieve high responsivity and signal-to-noise ratio. This level of functional partitioning and optimization is a breakthrough that was not possible with previous monolithic device designs. The resulting performance and scalability open up exciting new possibilities for ultra-compact, high-sensitivity photodetectors, imagers, and other optoelectronic integrated circuits.

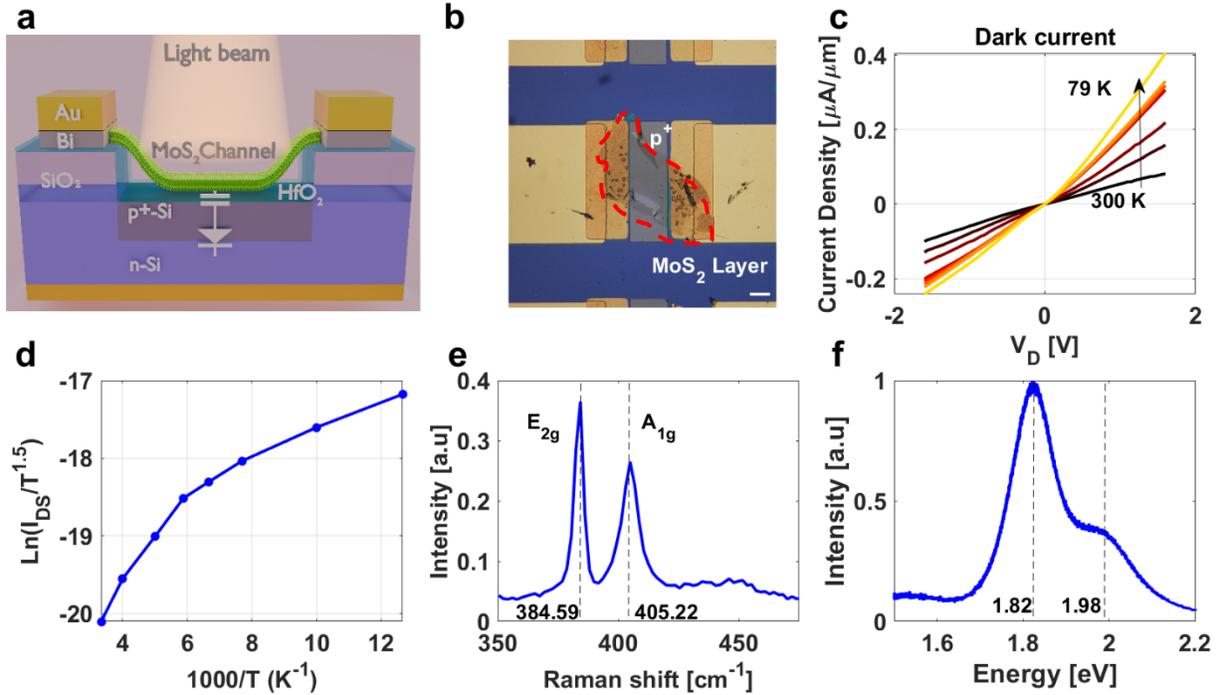

Figure 1. (a) 3D schematic of the device and (b) Optical microscopy image of a fabricated device. The reference bar is 20µm. (c) The current density versus drain-source voltage (Output characteristic) of the device at $V_g$=0V. (d) Arrhenius plot of the dark current at a fixed drain-source voltage of 1.6 V at $V_g$=0V. (e) Raman spectroscopy of $MoS_2$. (f) Photoluminescence spectrum of $MoS_2$ under green laser excitation ($\lambda$ = 532 nm).

Figure 1a shows a 3D schematic of our device, which consists of an $MoS_2$ flake (few layers thick) on top of an Si PN junction diode. A very thin $HfO_2$ layer (2 nm thick) formed on the PN junction isolates it electrically from the $MoS_2$ flake. This layer is deposited on the Si by atomic layer deposition prior to the $MoS_2$ flake placement. Under light illumination, the floating PN junction diode creates a photovoltage that gates the $MoS_2$ flake via the $HfO_2$ gate capacitor, thereby producing a photocurrent in the flake, which has Ohmic contacts. Fig.1b is an optical microscope image of a typical device, where fabrication has been completed. The edges of the $MoS_2$ flake are denoted by a red dashed line. The gray block is the $p^+$ doped Si region. Au/Bi electrodes (100

nm/20 nm thick) are in contact with the MoS$_2$ layer. They make Ohmic contacts because the work function of Bi, which is a semi-metal, is close to the conduction band minimum of MoS$_2$, thereby suppressing metal-induced gap states [6], [16]. Indeed, the current vs voltage (IV) characteristics are linear, as shown in Fig.1c. To confirm the Ohmic nature of the contacts, we measure the IV characteristics of the MoS$_2$ channel at cryotemperature. We find that these characteristics become slightly non-linear at a temperature lower than 170K. An Arrhenius plot of the current at a fixed bias is shown in Figure 1d. The positive slope observed in the Arrhenius plot confirms that the contact between the electrodes and the few layer MoS$_2$ channel is Ohmic [6], [16]. To assess the material quality of the MoS$_2$ flakes, which are exfoliated by the gold-mediated approach [17], Raman and photoluminescence (PL) spectroscopy are performed on the MoS$_2$ flake sitting on the 300 nm SiO$_2$. The Raman spectrum, shown in Figure 1e, exhibits two distinct peaks at 384.59 cm$^{-1}$ and 404.22 cm$^{-1}$, corresponding to the E$_{2g}$ and A$_{1g}$ vibrational modes of MoS$_2$, respectively. The separation between these two peaks is 20.63 cm$^{-1}$, indicating that the flake is few layer MoS$_2$. The PL spectrum, measured with green laser excitation ($\lambda$ = 532 nm), is shown as Figure 1f. It displays a strong peak at 1.82 eV due to the MoS$_2$ crystal and a weak peak around 1.98 eV that is caused by impurities or defects such as sulfur vacancies [18], [19].

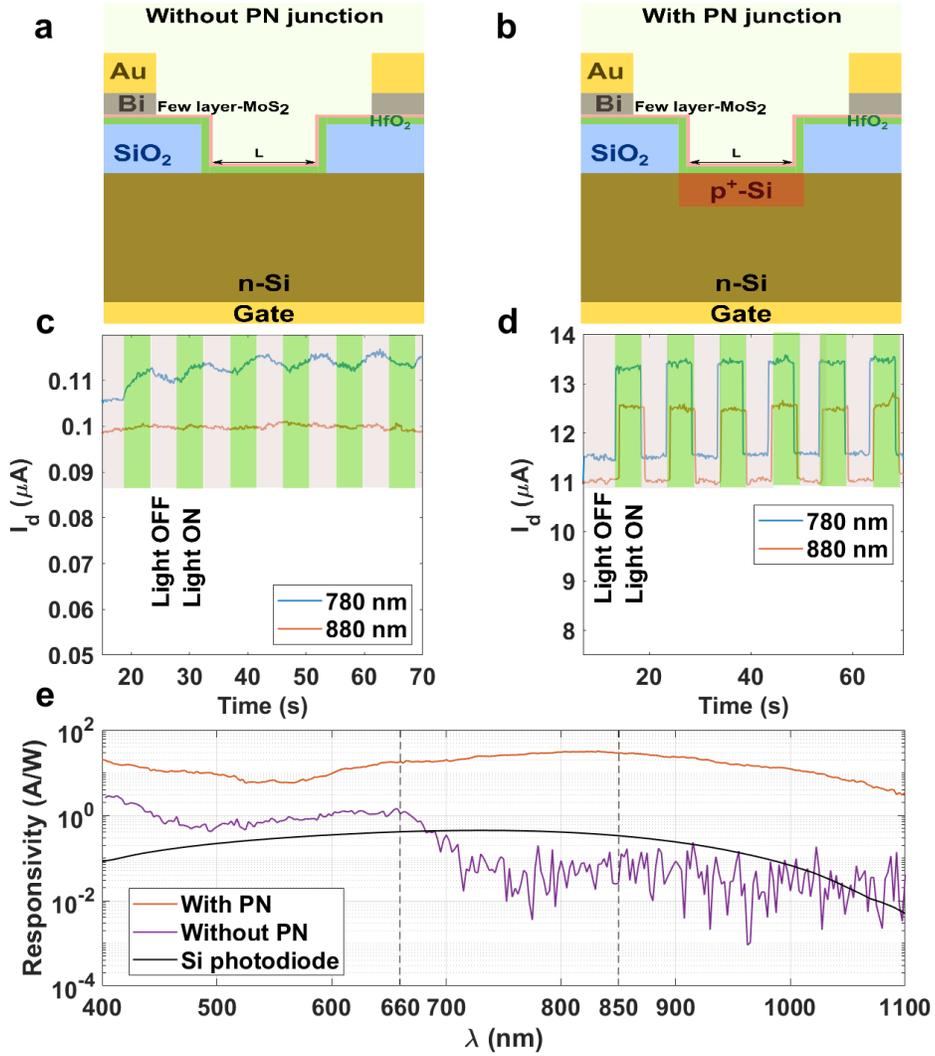

Figure 2. (a) Schematic of the device without a PN junction (WOJ). (b) Schematic of the device with a PN junction (WJ). (c) Current transients of the WOJ device at a bias of 1.6 V, where illumination at λ = 780 nm (blue curve) or λ = 880 nm (red curve) is pulsed ON/OFF periodically. (d) Current transients of the WOJ device at a bias of 1.6 V, where illumination at λ = 780 nm (blue curve) or λ = 880 nm (red curve) is pulsed ON/OFF periodically. (e) Photoresponsivity in logarithmic scale of the two devices and a commercial Si photodiode for comparison.

To elucidate the role of the PN junction below the MoS$_2$ channel, we fabricate two devices for comparison, one without a PN junction below the channel (denoted 'WOJ', meaning without junction, Figure 2a) and the second one with a PN junction (denoted 'WJ', meaning with junction, Figure 2b). The devices are biased at a fixed drain-source voltage of 1.6 V and the substrate is grounded. Light emitting diodes (λ= 780 nm and 880 nm, separately) are turned ON/OFF periodically to illuminate the devices. The WOJ device exhibits some noticeable photoresponse to the 780 nm illumination and no detectable photoresponse to the 880nm excitation as shown in Figure 2c. In contrast, the WJ device shows strong photoresponse for both wavelengths Figure 2d. Clearly, the strong photoresponses come from the photovoltage gating effect of the PN junction in the Si substrate.

To further investigate the underlying mechanism for the observed photoresponse, we measure the photoresponsivity of the two devices Figure 2e. According to the PL spectrum in Figure 1f, the bandgap of $MoS_2$ is 1.82eV, corresponding to an optical absorption edge at ~ 660 nm. It can be seen from Figure 2e (purple curve) that the photoresponse of the WOJ device is much stronger for wavelengths shorter than 660 nm. We attribute this to the fact that this spectral range is within the absorption edge of MoS2, i.e. where it absorbs light strongly. The photoresponse beyond the absorption edge (longer than 660nm) exponentially drops from 1 A/W at 660 nm to a level ($3 \times 10^{-2}$ A/W) that is limited by the device noise. It is for this reason that the WOJ device has somewhat weak photoresponse at 780 nm and clearly no detectable response at 880nm. For the WJ device, the photoresponsivity beyond the absorption edge reaches a maximum of 32 A/W at a wavelength of 832nm (460 times higher than WOJ), then it drops to 3.68 A/W at a wavelength of 1100 nm, in stark contrast with the WOJ device. The profile of spectral photoresponse follows a pattern similar to that of a silicon PN junction diode (black solid line in Fig.2e). Clearly, the strong photoresponse of the WJ device beyond the absorption edge comes from the photogating effect of the PN junction underneath. The spectral photoresponse within the absorption edge (< 660 nm) is similar to, but stronger by one order of magnitude than, that of the WOJ device. This is likely the combined effect of the photo absorption by MoS$_2$ and the photogating of the PN junction.

The photogating effect of the PN junction comes from the open circuit voltage V$_{oc}$ (photovoltage) under light illumination acting as a gate voltage that modulates the Fermi level of MoS$_2$ channel. It is not difficult to find that $V_{oc}$ is dependent on the incident light intensity $P_{light}$ following eq. (1) (**see SI Section I for derivation**).

$$V_{oc} = V_{th}\ln\left(\frac{P_{light}}{P_{light}^s} + 1\right) \qquad (1)$$

, where $V_{th} = \frac{\eta K_B T}{q}$ and $\eta$ is the ideality factor, $K_B$ the Boltzmann constant, T the absolute temperature and q the unit of charge. Note that the critical light intensity is given as $P_{light}^s = \frac{\hbar\omega J_s}{q\alpha} / \left(1 - e^{-\frac{W_{dep}}{l_a}}\right)$ (**see SI Section I for derivation**) where $\alpha$ denotes the transmittance of light incident on Si, $\hbar\omega$ the incident photon energy, $l_a$ the light absorption length and $W_{dep}$ the depletion region width of the PN junction. $J_s$ is the leakage current density of the PN junction and can be written as $J_s = \frac{qn_i W_{dep}}{2\tau_0}$ with $n_i$ being the intrinsic electron concentration of silicon substrate and $\tau_0$ the effective minority carrier recombination lifetime [20].

The photo induced open circuit voltage $V_{oc}$ supplies the gate voltage for the MoS$_2$ channel. A gate voltage $V_G$ applied on a 2D atomically thin semiconductor is related to the electron and hole concentrations (n and p) according to eq.(2a) [21]. The Fermi level $E_F$ shows up on the right side of the equation due to the quantum capacitance effect in atomically thin monolayers [22]. Note that $E_F$ in eq.(2a) is with respect to $E_i$, the middle energy level of the bandgap. $V_{mid}$ is the gate voltage needed to tune the Fermi level to the middle of the bandgap ($E_F = 0$).

$$q(V_G - V_{mid}) = E_F + \frac{q^2(n-p)}{C_{ox}} \qquad (2a)$$

, where $C_{ox}$ is the gate capacitance per unit area.

The shift of the gate voltage by $V_{mid}$ is caused by charged trap states, which can be described by eq.(2b).

$$V_{mid} = \frac{Q_{ss}}{C_{ox}} = \frac{1}{C_{ox}} \int_{E_v}^{E_F} qD_{it}\, dE \qquad (2b)$$

, where $Q_{ss}$ is the surface trapped charges per unit area and $D_{it}$ is the density of interface trap states in the bandgap. We assume that the trap states below $E_F$ are charged by filling with electrons.

The electron n and hole concentration p (cm$^{-2}$) in the 2D monolayer semiconductor are related to the Fermi level $E_F$ in the same way as for 3D semiconductors, as given in eq.(2c) [23], [24].

$$n = n_{i2D} e^{\frac{E_F}{K_B T}}, \quad p = n_{i2D} e^{-\frac{E_F}{K_B T}} \qquad (2c)$$

, where $n_{i2D}$ is the electron concentration per unit area of intrinsic MoS$_2$ 2D monolayers.

Under light illumination, the open circuit voltage $V_{oc}$ of the PN junction under illumination shifts the gate voltage, so we write $V_{oc} = \Delta V_G$. This change in gate voltage in turn results in changes to $V_{mid}$, $E_F$ and $n$ and $p$ in MoS$_2$ as shown in eq.(3a). It is not difficult to find that $\Delta V_{mid}$, $\Delta n$ and $\Delta p$ are related with $\Delta E_F$ following eqs.(3b) and (3c) from the derivatives of eqs.(2b) and (2c), respectively.

$$q(\Delta V_G - \Delta V_{mid}) = \Delta E_F + \frac{q^2(\Delta n - \Delta p)}{C_{ox}} \qquad (3a)$$

$$\Delta V_{mid} = \frac{\Delta Q_{ss}}{C_{ox}} = \frac{qD_{it}\Delta E_F}{C_{ox}} \qquad (3b)$$

$$\Delta n = \frac{n}{K_B T}\Delta E_F, \quad \Delta p = -\frac{p}{K_B T}\Delta E_F \qquad (3c)$$

The photocurrent in the MoS$_2$ channel is given by Ohm's Law and can be expressed as eq.(4a) by plugging eqs.(1) and properly reformatting eqs.(3a-c).

$$I_{ph} = q(\mu_n \Delta n + \mu_p \Delta p)V_{ds}\frac{W}{L} = I_{th}\ln\left(\frac{P_{light}}{P^S_{light}} + 1\right) \qquad (4a)$$

, where L and W are the length and width of MoS$_2$ channel gated by the PN junction, respectively, and $I_{th}$ is defined as the threshold current for convenience which is given by $I_{th} = \eta q(\mu_n n - \mu_p p)V_{ds}\frac{W}{L}/\left(\frac{q^2 D_{it}}{C_{ox}} + 1 + \frac{q^2(n+p)}{kTC_{ox}}\right)$. Note that the denominator in the parentheses can be tailored close to zero if the density of defects is minimized and a negative capacitance is adopted by using ferroelectric materials as the gate dielectrics[25], [26], [27]. In this case, highly sensitive photodetectors can be potentially built.

The photoresponsivity can be written as follows:

$$R = \frac{I_{ph}/q}{P_{light} \times A_c/\hbar\omega} = R_{max}\frac{P^S_{light}}{P_{light}}\ln\left(\frac{P_{light}}{P^S_{light}} + 1\right) \qquad (4b)$$

, where $A_c$ is the cross-sectional area of the PN junction on a plane that is transverse to the direction of the incident light, $R_{max}$ is the maximum photoresponsivity in the limit that the illumination intensity approaches zero, and is given by $R_{max} = \frac{I_{th}/q}{P^S_{light} \times A_c/\hbar\omega}$.

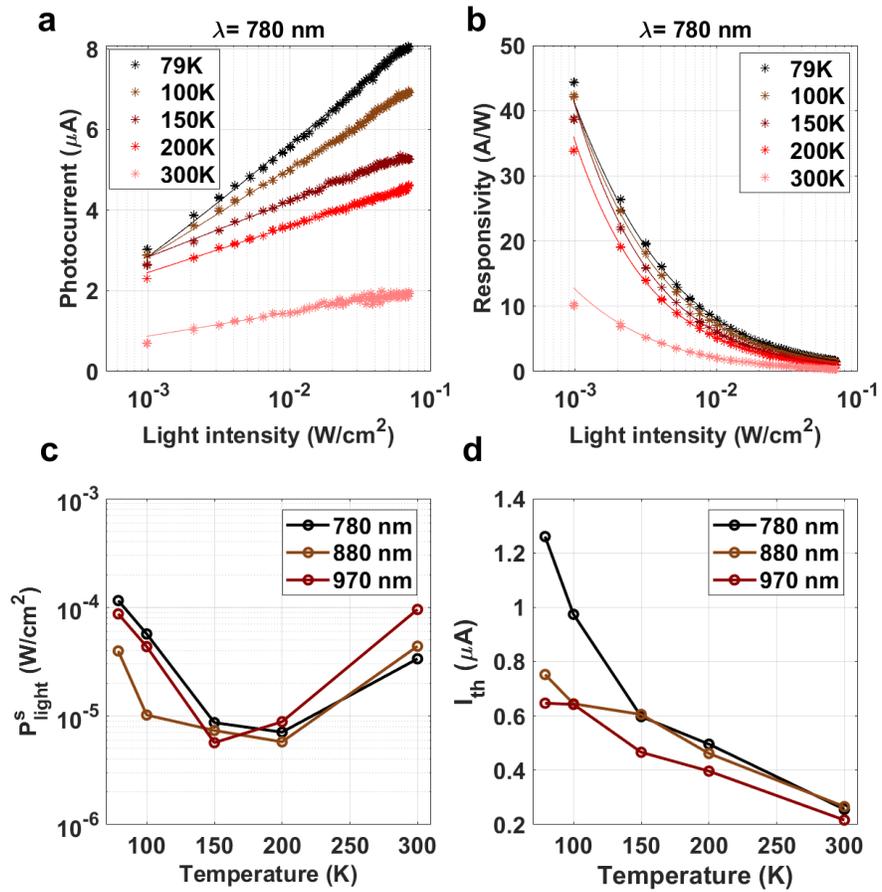

Figure 3. (a) Measured photocurrent vs illumination intensity, for illumination at $\lambda = 780$ nm and at different temperatures. (b) Measured responsivity vs illumination intensity, for illumination at $\lambda = 780$ nm and at different temperatures. Asterisks are measured data, while curves are from our model. (c) Extracted critical light intensity as a function of temperature, for illumination at different wavelengths. (d) Extract threshold current as a function of temperature for illumination at different wavelengths.

To investigate that the photoresponses of our photodetector are governed by eqs. (4a-b), we measure device response as a function of illumination wavelength ($\lambda = 780$nm, 880nm, 970nm) and as a function of temperature. In this case, the light absorption by MoS$_2$ is negligible and photoresponses solely come from the photogating effect of the PN junction in the Si substrate. The measured photocurrent vs illumination intensity (from $10^{-3}$ to $10^{-1}$ Wcm$^{-2}$, λ = 780 nm) at different temperatures (79 K, 100 K, 150 K, 200 K, and 300 K) is shown as Figure 3a. From this data, we determine the responsivity, and present the results as Figure 3b. (We set the light intensity to a relatively high value due to the presence of non-ideal effects. These non-ideal effects include the influence of surface states around the MoS$_2$ material and the contact resistance between the substrate and the ground electrode. These non-ideal factors compromise the photo-induced open-circuit voltage and its ability to effectively gate the device.) In Figure 3a and 3b, the predictions of our model (Eq. 4a and 4b) are also plotted, where the fitting parameters are the critical light intensity $P_{light}^s$ and the threshold current $I_{th}$. The critical light intensity $P_{light}^s$ indicates the performance of the PN junction as a photovoltage generator upon light illumination. The threshold current $I_{th}$ is the parameter showing how the MoS2 flake respond in current to the photogate voltage. These parameters are temperature dependent and plotted in Figure 3c and d. $P_{light}^s$ first declines as temperature decreases from room temperature due to the fact that the intrinsic electron concentration $n_i$ of semiconductors goes low as temperature decreases. $P_{light}^s$ then increases after temperature is lower than ~ 170K. At a lower temperature, the device becomes more sensitive to illumination at lower light intensity. The relatively high illumination intensity will saturate the PN junction, reducing the sensitivity of open circuit voltage upon light illumination. It reflects in the higher value of $P_{light}^s$ at lower temperature.

Now let us take a look at the threshold current $I_{th}$. For our device, MoS$_2$ is n-type and the electrons are the majority carriers. The minority holes in the expression of $I_{th}$ can be neglected. In this case, we have $I_{th} = \eta I_{dark}/(\frac{q^2 D_{it}}{C_{ox}} + 1 + \frac{q^2 n}{kTC_{ox}})$ with the dark current $I_{dark}$ in MoS2 given by $I_{dark} = q\mu_n n V_{ds} \frac{W}{L}$. From Fig. 1c, it can be seen that the dark current increases as temperature decreases. The threshold current $I_{th}$ should therefore increase with decreasing temperature because it is proportional to the dark current, provided that the other terms (e.g. $\frac{q^2 n}{kTC_{ox}}$ in denominator) play a minor role. It can be seen from Figure 3d that the measured threshold current does indeed increase monotonically with decreasing temperature.

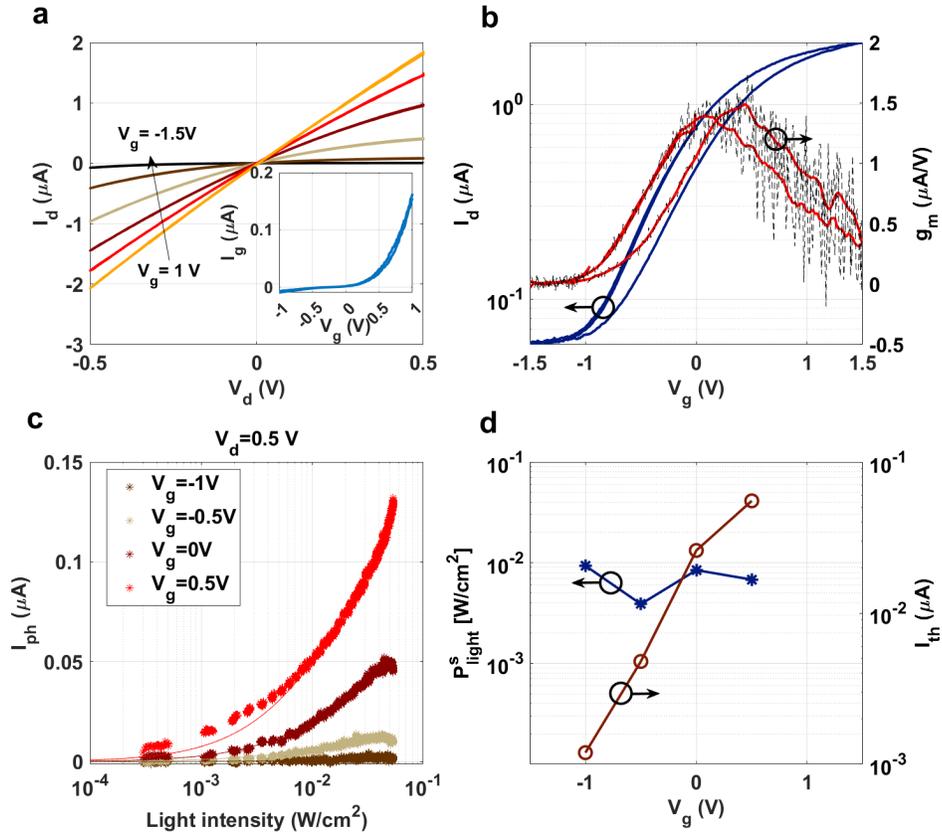

Figure 4. (a) IV curve at different gate voltages. The inset shows the gate current versus gate voltage with $V_{ds}$ = -0.5V. (b) Gate transfer characteristics and transconductance of MoS2 phototransistor. The drain-source voltage is fixed at 0.5V. (c) Photocurrent vs light intensity at different gate voltages. (d) Extracted critical light intensity and threshold current as a function of gate voltage.

We next investigate the role of the gate in the phototransistor performance by measuring the terminal currents as a function of gate bias. Since the leakage current through the 2 nm thick $HfO_2$ layer is smaller than the leakage current through the PN junction (**see SI Section II for leakage current vs device current**), the externally applied gate voltage on the substrate will mostly drop on the $HfO_2$ capacitor unless $V_g > 0.5V$. It means that the existence of the PN junction does not affect the substrate serving as a back gate. For this reason, we directly applied the gate voltage on the substrate. The IV curves of the $MoS_2$ device under different gate voltage are shown in Fig.4a. The drain-source voltage is fixed at 0.5V. The IV curves are linear at positive gate voltage but become increasingly nonlinear at more negative gate voltage. This is because the $MoS_2$ channel starts to be pinched off at negative gate voltages, creating a potential barrier in the $MoS_2$ channel between source and drain. The source-drain bias will slightly modulate the profile of the potential barrier, inducing some nonlinearity in the IV curve.

It is interesting to note that the gate leakage is asymmetric at negative and positive gate voltage (inset in Fig.4a). It is likely because the $MoS2$-$HfO_2$-$Si(p^+)$ structure is in depletion mode for $V_g < 0V$ and accumulation mode for $V_g > 0V$. In the depletion mode, part of the applied gate voltage will drop on the depletion region and the actual voltage drop on the 2 nm thick $HfO_2$ will be smaller, resulting in a small leakage current. In contrast, in the accumulation mode, all of the applied gate voltage will directly drop on the $HfO_2$, resulting in a significantly higher leakage current.

Figure 4b displays the gate transfer characteristics of the same device, along with the transconductance found from the experimental data using the formula $g_m = \frac{dI_d}{dV_g}$ on the right y-axis. The gray dashed line along the transconductance curve is the actual transconductance and the solid red line is the smoothed curve of the transconductance The maximum transconductance is observed at 0 or 0.5V depending on the sweep direction due to the defect-induced hysteresis in gate transfer characteristics, which is consistent with the fact that the measured photocurrent is maximized at a gate voltage of $V_G = 0.5V$ (Figure 4c). This observation is in good agreement with our theory in which the photo-induced open circuit voltage of the PN junction serves the gate voltage. A higher transconductance means a higher photoresponse.

Figure 4d presents $P_{light}^S$ extracted from Figure 4c by fitting the experimental data with eq.(4a). According to our theory, $P_{light}^S$ should be independent of the gate voltage. Indeed, as shown in Figure 4d left y-axis, $P_{light}^S$ remains relatively constant at different gate bias. As discussed, we have $I_{th} = \eta I_{dark}/(\frac{q^2 D_{it}}{C_{ox}} + 1 + \frac{q^2 n}{kTC_{ox}})$. A negative gate voltage decreases the dark current by nearly two orders of magnitude (see Fig.4b). The existence of defect states will make the denominator of $I_{th}$ less dominated by the electron concentration term, resulting in a proportional decrease of $I_{th}$ as the gate voltage goes negative. At negative gate voltage, the appearance of surface depletion region will reduce the effective gate capacitance (dielectric capacitance $C_{ox}$ is in series with the surface depletion capacitance), which will further reduce $I_{th}$. This explains why $I_{th}$ drops by nearly 3 orders of magnitude as the gate voltage sweeps to -1 V, one order of magnitude more than the dark current decreases.

**Conclusion**:

In this work, we have demonstrated a novel MoS₂ phototransistor in which the photovoltage of the PN junction in the substrate acts as a backgate voltage, inducing a high photoresponse in MoS₂. Our formulated photoresponse theory provides a coherent framework that aligns well with the experimental observations for this phototransistor configuration. This unique device structure separates the light absorption and electrical response, providing us an opportunity to design new photodetectors with high performance. For example, further enhanced photoresponses are possible when a negative gate capacitance is introduced by using ferroelectric materials, because a negative gate capacitance can potentially reduce or even zero out the denominator in the expression of threshold current $I_{th}$.

**Experimental**

**Device Fabrication** The detailed fabrication process is shown in **SI Section III**. The fabrication process starts with a n-type silicon substrate with a background resistivity of $0.1 - 0.01\ \Omega cm^{-1}$, which corresponds to a doping concentration around $10^{17} cm^{-3}$. Using standard photolithography techniques, a rectangular region of size $L \times 200\ \mu m$ is opened in the 300 nm thick oxide layer on the substrate, using a wet etching process with buffered hydrofluoric acid (B-HF). For our devices L varies from 10μm to 60μm. Next, ion implantation is performed to introduce a 150 nm thick boron-doped layer with a concentration of $10^{19} cm^{-3}$ in the etched SiO₂ region. The sample is then cleaned using acetone and isopropyl alcohol (IPA), and a rapid thermal annealing (RTA) step is carried out to activate the dopants. Then we washed the sample again with the same procedure and prepared it for Atomic Layer Deposition (ALD). Following this, a 2 nm thick layer of hafnium dioxide (HfO₂) is deposited on the sample using ALD. The sample is then washed with Ethanol, IPA then water to prepare it for the transfer of MoS₂. The MoS₂ is transferred onto the sample using a gold-mediated exfoliation technique, as described in reference [17]. After the transfer, the sample is annealed for 1 hour at 420 K in vacuum (pressure: 1×10$^{-2}$ Pa). This annealing is necessary for releasing any tension or water molecules on the MoS$_2$. Finally, to ensure ohmic contact between the electrodes and the MoS$_2$ layer, standard photolithography is performed, followed by the thermal evaporation of a 20 nm thick bismuth layer+100nm gold layer, which forms an ohmic contact with the MoS$_2$ [6], [16].

**Electrical Characterization** All measurements are conducted under vacuum (pressure: $10^{-2}$ Pa). The sample was first annealed at 420 K for 1 hour and then allowed to cool down for an additional 1 hour under vacuum. Except for the responsivity measurements in different wavelengths shown in Figure 2e which we used a portable cryostat LANHAI science instrument, we used closed "Cycle Cryogenic Probe Station-LH-CRPS-5K" from LANHAI science instrument for all other measurements. The LEDs used in all experiments were M780L3, M880L3, M970L4, and M1050L4 from Thorlabs. An ACL2520U-B lens was used for collimation.

**Photoresponse Measurements** We used the modulation mode of the LEDD1B (Thorlabs) LED driver. We employed a function generator to produce a square wave with a base frequency of 100 mHz and a linear amplitude modulation from 2 V to 0 V over 10 minutes. This voltage was applied to the modulation input of the LED driver. We then measured the device current under both light and dark conditions using a commercial InGaAs PIN photodiode (G10899-3K) from Hamamatsu, as well as our own device. The current measurements were performed using a 2636B Keithley Source meter, applying a constant voltage and measuring the current during the measurement period. We wrote a LabVIEW VI to synchronize and control all the instruments together. Finally, we subtracted the device current under dark conditions from the device current under light conditions. We calculated the light intensity using the responsivity curve reported in the datasheet of the G10899-3K photodiode, and then assigned the specific intensity to the corresponding time to plot the photocurrent versus light intensity, as shown in Figure 3a, b and Figure 4c.

For responsivity versus wavelength measurements, as shown in Figure 2e, we employed a custom-built setup at the side exit of the iHR320 monochromator. This setup utilized a portable cryostat for both our device and a commercial photodetector. The monochromator was equipped with an LSH-T250 tungsten broadband light source and three gratings blazed at 350 nm (grating 1), 900 nm (grating 2), and 1500 nm (grating 3). In our measurements, we used gratings 1 and 2, both with a groove density of 1200 gr/mm. We precisely controlled the incident wavelength by setting the exit slit-width to 1 mm, which corresponds to a 2.31 nm bandpass for the 1200 gr/mm grating. This allowed us to systematically evaluate the photodetector's response across different spectral ranges from 400 nm to 700 nm. For longer wavelengths (700 nm to 1400 nm), we implemented two modifications to the setup. First, we introduced an FGL715 (Thorlabs) filter into the beam path to block unwanted high-frequency components generated by the monochromator. Second, we switched to the monochromator's second grating, which was optimized for 900 nm. To measure the current while applying voltage to the device, we used a 2636B Keithley SourceMeter. As a reference photodetector, we employed the same InGaAs PIN photodiode housed within a portable cryostat from LANHAI science instrument. To measure both dark current and light current, we developed a virtual instrument (VI) in LabVIEW to control both the SMU2636B and the iHR320. This VI closed the slit to measure the dark current and opened the slit to 1 mm to measure the device current under illumination. Finally, we calculated the photocurrent by subtracting the device current under light conditions from the current under dark conditions. The light intensity calculation is again similar to the previous section.

**Author Contributions**

Y.D. conceived the idea and supervised the research. SS. M and Y.D wrote the manuscript. SS.M. designed the experiments setup for characterizations, upgraded the LabVIEW codes for data collection, performed the measurements. SS.M, Y.D, J.W. and K. C. analyzed and discussed the data. J.W., Y.L., Z. F. and K. L. helped SS.M. with the device fabrication. Y.L. did the PL and Raman spectroscopy. Z. F. did the ion implantation and RTA for the samples. All authors discussed the results and commented on the manuscript.

**Acknowledgement**

This work was financially supported by Oceanic Interdisciplinary Program of Shanghai Jiao Tong University (SL2022ZD107), Shanghai Pujiang Program (22PJ1408200), National Science Foundation of China (NSFC, No. 92065103), and Shanghai Jiao Tong University Scientific and Technological Innovation Funds (2020QY05). The devices were fabricated at the center for Advanced Electronic Materials and Devices (AEMD), and Raman spectrum and photoluminescence measurements were carried out at the Instrumental Analytical Center (IAC), Shanghai Jiao Tong University. This work was supported in part by the Australian Research Council Centre of Excellence for Transformative Meta-Optical Systems (Project No. CE200100010). SS.M. acknowledges a scholarship from the University of Melbourne – Shanghai Jiaotong University (SJTU) Joint PhD program.